\documentclass[aip,apl,letterpaper,amsmath,amssymb,superscriptaddress,reprint]{revtex4-1}
\usepackage{graphicx}
\usepackage{microtype}
\usepackage{bm}
\usepackage{hyperref}

\begin{document}

\title{Influence of the Hall-bar geometry on harmonic Hall voltage measurements of spin-orbit torques}
\author{Lukas Neumann}
\author{Markus Meinert}
\email{meinert@physik.uni-bielefeld.de}
\affiliation{Center for Spinelectronic Materials and Devices, Department of Physics, Bielefeld University, D-33501 Bielefeld, Germany}

\date{\today}

\begin{abstract}
Harmonic Hall voltage measurements are a wide-spread quantitative technique for the measurement of spin-orbit induced effective fields in heavy-metal / ferromagnet heterostructures. In the vicinity of the voltage pickup lines in the Hall bar, the current is inhomogeneous, which leads to a hitherto not quantified reduction of the effective fields and derived quantities, such as the spin Hall angle or the spin Hall conductivity. Here we present a thorough analysis of the influence of the aspect ratio of the voltage pickup lines to current channel widths on the apparent spin Hall angle. Experiments were performed with Hall bars with a broad range of aspect ratios and a substantial reduction of the apparent spin Hall angle is already seen in Hall crosses with an aspect ratio of 1:1. Our experimental results are confirmed by finite-element simulations of the current flow.
\end{abstract}

\maketitle

\section{Introduction}

The spin Hall effect \cite{Dyakonov1971, Hirsch1999, Hoffmann2013, Sinova2015} converts a charge current density $j$ into a transverse spin current density $j_\mathrm{s}$. The charge-to-spin conversion efficiency is characterized by the spin Hall angle (SHA) $\theta_\mathrm{SH} = j_\mathrm{s} / j$. The spin Hall angle of crystalline materials is in many aspects experimentally\cite{Qiu2013, Sagasta2016, Nguyen2016, Schulz2016, Zhang2017} and theoretically\cite{Tanaka2008, Lowitzer2011, Freimuth2010, Gradhand2012, Koedderitzsch2015} well understood and various heavy metals (HM) with large spin Hall angle were identified, such as Pt\cite{Sagasta2016}, $\beta$-W \cite{Pai2012}, $\beta$-Ta \cite{Liu2012a}

\begin{figure}[b]
\includegraphics[width=8.6cm]{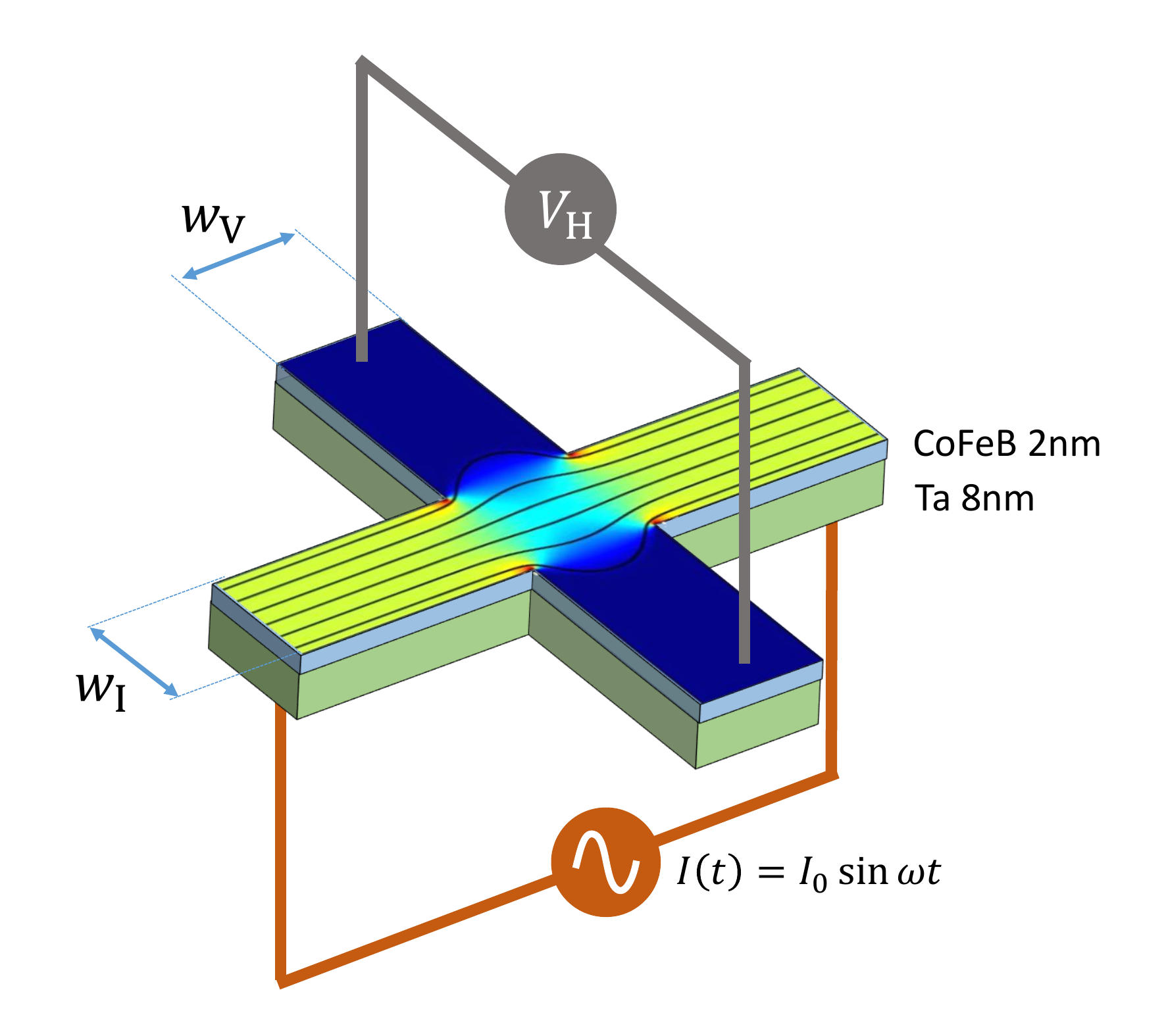}
\caption{\label{fig:HallCross} Schematic of the experimental setup. The current line width $w_\mathrm{I}$ and voltage pickup line width $w_\mathrm{V}$ are indicated. The current density profile was obtained from a finite-element simulation.}
\end{figure}

The harmonic Hall voltage measurement technique has become a standard technique to determine the magnitude of the so-called spin-orbit torques (SOT) or effective fields originating from the spin current flowing into an adjacent ferromagnetic (FM) layer and allows for a quantitative determination of $\theta_\mathrm{SH}$.\cite{Pi2010, Garello2013, Kim2013, Avci2014a, Hayashi2014, Avci2014b, Qiu2014, Wen2017, Lau2017, Yun2017} For this method, Hall-bar structures are patterned into HM / FM bilayers and ac currents  $I(t) = I_0 \sin \omega t$ are driven through the current channels. As a consequence of the current-dependent effective fields that lead to an in-phase modulation of the magnetization orientation, the resulting Hall voltage $V_\mathrm{H}$ has first and second harmonic components
\begin{equation}
V_\mathrm{H}(t) = R_\mathrm{H} I_0 \sin(\omega t) + R_\mathrm{H}^{2\omega} I_0 \cos(2\omega t),
\end{equation}
which can be measured by Fourier transformation of a time series or, more commonly, by a lock-in amplifier. Depending on the orientation of the magnetization, various analytical expressions were proposed for the analysis of the second harmonic Hall voltage, which rely on small-angle approximations of the modulation angles. The second harmonic Hall resistance $R_\mathrm{H}^{2\omega}$ is proportional to the current density, such that $V_{2\omega} = R_\mathrm{H}^{2\omega} I_0 \propto j^2$, whereas $V_\omega = R_\mathrm{H}I_0 \propto j$. Therefore, the dc Hall voltage and $V_\omega$ do not depend on the width of the voltage pickup lines of the Hall bar. However, due to its quadratic current density dependence, one may expect a reduction of $V_{2\omega}$ when the voltage pickup width is large and the current density becomes inhomogeneous in the Hall bar. Here, we systematically study V$_{2\omega}$ and the corresponding apparent spin Hall angle for Hall bars with different aspect ratios $a = w_\mathrm{V} / w_\mathrm{I}$ of the current line widths $w_\mathrm{I}$ and voltage pickup line widths $w_\mathrm{V}$ as depicted in Fig. \ref{fig:HallCross} in $\beta$-Ta / CoFeB bilayer structures with in-plane magnetic anisotropy.

\section{Experiment}

A thin film heterostructure of Si / SiO$_2$~50\,nm / Ta~8\,nm / Co$_{40}$Fe$_{40}$B$_{20}$~2\,nm / MgO~1.8\,nm / Ta~1.5\,nm was grown by dc and rf magnetron sputtering. Hall bars were written by electron beam lithography and ion beam milling. In all cases, the current channel width $w_\mathrm{I} = 15\,\mathrm{\mu m}$ was kept fixed, while the voltage pickup line width $w_\mathrm{V}$ was varied between 1\,$\mu$m and 40\,$\mu$m. Harmonic Hall measurements were performed by wire bonding and mounting the samples in a vector magnet. An ac current $I(t) = I_0 \sin \omega t$ was driven through the current channel such that the current density was $j = 2\times 10^{10}\,\mathrm{A/m^2}$. The first and second harmonic components of the Hall voltage were simultaneously detected with a multi-demodulator lock-in amplifier (Zurich Instruments MFLI-MD) at $f = \omega / 2\pi = 3121\,\mathrm{Hz}$. The second harmonic out-of-phase Hall voltage amplitude $V_{2\omega}$ can be written as\cite{Wen2017,Avci2014b, Note}
\begin{equation}\label{eq:harmonichall}
V_{2\omega} =  \left( - \frac{B_\mathrm{FL}}{B_\mathrm{ext}} R_\mathrm{P} \cos 2 \varphi 
 - \frac{1}{2} \frac{B_\mathrm{DL}}{B_\mathrm{eff}} R_\mathrm{A} + \alpha' I_0 \right) I_0 \cos \varphi.
\end{equation}
The angle $\varphi$ is the angle between current and magnetization and $B_\mathrm{eff} = B_\mathrm{ext} + B_\mathrm{sat}$ is the effective field. The out-of-plane saturation field $B_\mathrm{sat} = B_\mathrm{dem} - B_\mathrm{ani}$ and the anomalous Hall resistance amplitude $R_A$ were obtained from Hall voltage measurements in a perpendicular magnetic field up to 2.2\,T. We found $B_\mathrm{sat} = 0.64\,\mathrm{T}$ and $R_\mathrm{A} = 1.46\,\Omega$. The planar Hall amplitudes $R_\mathrm{P}$ were obtained from the first harmonic $V_\omega = R_\mathrm{P} I_0 \sin 2\varphi$. The term $\alpha' I_0$ describes a parasitic contribution arising from the anomalous Nernst effect (ANE), which yields an electric field $\bm{E}_\mathrm{ANE} = -\alpha \nabla T \times \bm{m} \propto I_0^2$. The prefactor $\alpha'$ summarizes all geometrical parameters and the film resistivity, heat conductivity, etc. that determine $\nabla T$. The above formula was fitted to the experimental data and damping-like effective fields and anomalous-Nernst contributions were separated by their dependence on the external field. The spin Hall angle was obtained from the damping-like effective field as
\begin{equation}
\theta_\mathrm{SH} = \frac{2e}{\hbar} \frac{ B_\mathrm{DL} M_\mathrm{s} t_\mathrm{CFB} }{j_0^\mathrm{Ta}},
\end{equation}
where $j_0^\mathrm{Ta}$ is the current density amplitude in the Ta layer far away from the Hall voltage pickup lines. The magnetization of the CoFeB film was determined by alternating gradient magnetometry to be $M_\mathrm{s} = (1050 \pm 50)\,\mathrm{kA/m}$. The resistivity of both the Ta and CoFeB layers was about $200 \times 10^{-8}\,\mathrm{\Omega m}$, such that no correction for current shunting due to unequal resistivities was required.

\begin{figure}
\includegraphics[width=8.6cm]{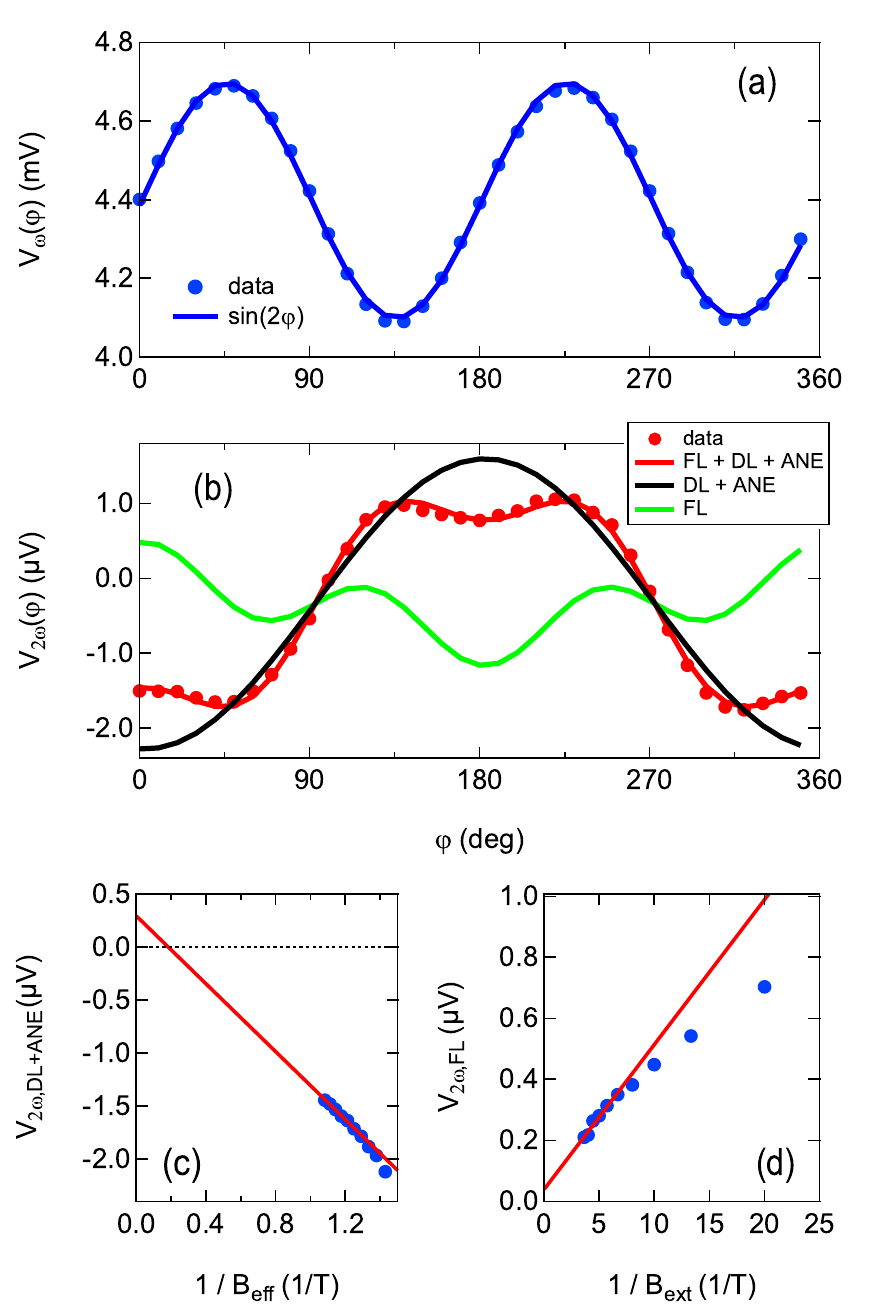}
\caption{\label{fig:measurements} (a), (b): Measurements of $V_\omega(\varphi)$ and $ V_{2\omega}(\varphi)$ on a Hall cross with $w_\mathrm{V} = 15\,\mathrm{\mu m}$ at 20\,mT and $j = 2 \times 10^{10}\,\mathrm{A/m^2}$. Fits are included as discussed in the main text. (c), (d):  $V_{2\omega,\mathrm{DL+ANE}}$ and $V_{2\omega,\mathrm{FL}}$ as discussed in the main text with line fits done on the interval $B_\mathrm{ext} \in [0.125\,\mathrm{T}, 0.275\,\mathrm{T}]$.}
\end{figure}

\begin{figure*}
\includegraphics[width=17cm]{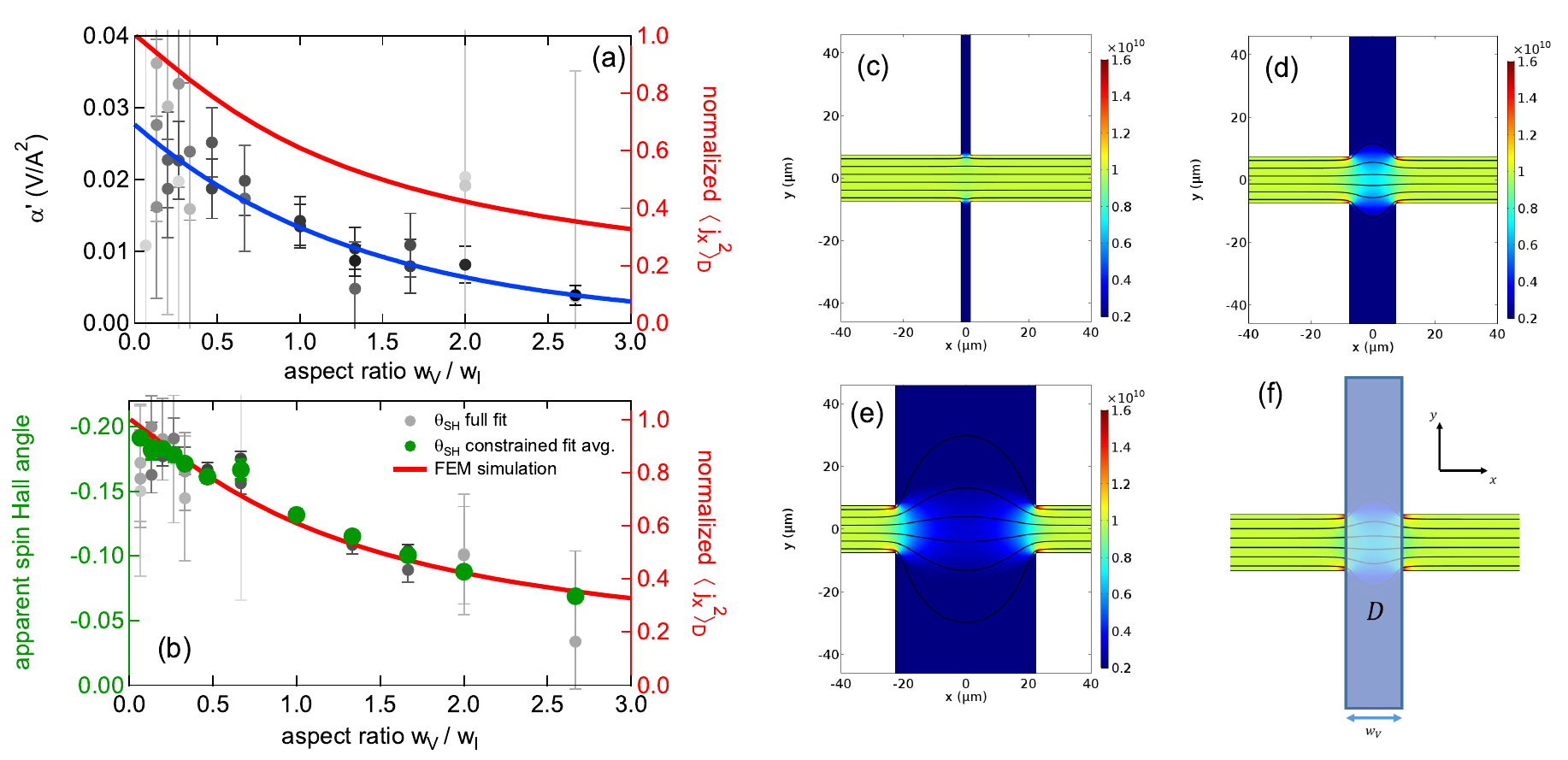}
\caption{\label{fig:ane_sha_fem} (a): Apparent anomalous Nernst parameter $\alpha'$ as a function of the Hall-bar aspect ratio $a = w_\mathrm{V} / w_\mathrm{I}$ with an exponential fit.  In addition, the normalized $\left< j_x^2 \right>_D$ from the finite-element simulations is shown. The experimental data points are color-coded, where lighter color represents larger fit error.  (b): Apparent spin Hall angle as a function of the Hall-bar aspect ratio. (c)-(e): Finite-element simulations of the current density in Hall crosses with aspect ratios of $a=0.2$ (c), $a=1$ (d), and $a=3$ (e). (f): Graphical representation of the domain $D$ over which the average of $j_x^2$ is taken. }
\end{figure*}

\section{Results}
In Fig. \ref{fig:measurements}(a),(b) we show a typical $V_\omega$, $V_{2\omega}$ measurement, which demonstrates the fitting procedure of the experimental $\varphi$-scans with Eq. \ref{eq:harmonichall} and verifies the presence of FL and (DL + ANE) contributions. In Fig. \ref{fig:measurements}(c),(d) we show the the dependence of the second harmonic Hall contributions  $V_{2\omega,\mathrm{DL+ANE}}$ and $V_{2\omega,\mathrm{FL}}$ on the effective field and the external field, respectively. The slopes of the line fits were used to determine the effective fields $B_\mathrm{FL}$ and $B_\mathrm{DL}$, the y-axis intercept relates to $\alpha'$. At small fields, deviations from the expected lines are observed, which arise from a uniaxial magnetic anisotropy of the films due to the sputtering process. A nonzero ANE is observed, as is indicated by the intercept of the linear fit in Fig. \ref{fig:measurements}(c) with the y-axis. The high-field line fit of the FL contribution extrapolates to nearly zero for $1/B_\mathrm{ext} \to 0$, which indicates that the magnetization is essentially saturated in the field range used for the fitting.

The anomalous Nernst parameter $\alpha'$ is shown in Fig. \ref{fig:ane_sha_fem}\,(a) as a function of the Hall-bar aspect ratio $a$. Although substantial scatter is present in the data, the reliable data points with small fit error (indicated by heavier color) show a clearly decreasing trend with increasing $a$. An empirical exponential weighted fit to these data was used to reduce the scatter on the measurement of $\theta_\mathrm{SH}$, which is shown in Fig. \ref{fig:ane_sha_fem}\,(b). In the plot, data denoted as ``$\theta_\mathrm{SH}$ full fit'' (light grey) were directly obtained from fits as shown in Fig. \ref{fig:measurements}\,(c). Large errors seen in some measurements arise from noisier $V_{2\omega}$ measurements. Data denoted as ``$\theta_\mathrm{SH}$ constrained fit avg.'' (dark green) were obtained as inverse-variance weighted averages over three Hall crosses per aspect ratio and using the empirical fit to $\alpha'$ in Fig. \ref{fig:ane_sha_fem}\,(a).  Remarkably, $\theta_\mathrm{SH}$ has a very similar decreasing trend as $a$ increases. Since both the ANE and the SOT contributions depend quadratically on the current density, a similar trend with respect to the aspect ratio is expected. Notably, a Hall cross with fourfold symmetry ($a = 1$) has an apparent spin Hall angle that is only $\approx 69\%$ of the true value as obtained in Hall bars with a small $a$. The spin hall angle approaches $\theta_\mathrm{SH} = -0.19$ for small $a$, which corresponds to a spin Hall conductivity of $\sigma_\mathrm{SH} = \theta_\mathrm{SH} / \rho_\mathrm{Ta} = -95\,000\,\mathrm{S/m}$. Our result on the symmetric Hall crosses ($\theta_\mathrm{SH}(a=1) \approx -0.13$) is in line with previous measurements using similar Hall cross structures.\cite{Qiu2014, Hao2015, Jamali2016} The ANE electric field at $a \approx 0$ is $E_{2\omega, \mathrm{ANE}} = V_{2\omega, \mathrm{ANE}} / w_\mathrm{I} = 0.034\,\mathrm{V/m}$. For better comparability, we normalize the result to $E_{2\omega, \mathrm{ANE}} / j_0^2 = 4.25 \times 10^{-23}\,\mathrm{Vm^3/A^2}$, which is slightly smaller than $E_{2\omega, \mathrm{ANE}} / j_0^2 = 5.3 \times 10^{-23}\,\mathrm{Vm^3/A^2}$ obtained by Avci et al. in Ta 6nm / Co 2.5nm stacks with $a = 0.5$.\cite{Avci2014b} The larger difference of resistivities in Ta / Co gives rise to a larger temperature gradient as compared to our Ta / CoFeB heterostructure with similar layer resistivities.

To gain a deeper understanding of the observed reduction of $\alpha'$ and $\theta_\mathrm{SH}$, we performed finite-element (FEM) simulations of the current density distributions. In Fig. \ref{fig:ane_sha_fem}\,(c)-(e) we show current density distributions in three different Hall bars with $a = 0.2 ,1, 3$. Only weak current leakage into the voltage pickup lines is seen when $a \approx 0$, while the current density clearly becomes strongly inhomogeneous when $a \approx 1$. In the extreme case of $a \gg 1$, the current density leaks strongly into the voltage pickup lines and is greatly reduced in the core region of the Hall bar. To understand the influence of the inhomogeneous current density on the measurement of the spin Hall angle, we remind that only the current component parallel to $x$ contributes to the measured Hall voltage. Therefore, the measured $V_{2\omega}$ can only depend on $j_x^2$.  In the finite-element simulations, we can directly access $j_x^2$ and compute the average $\left< j_x^2 \right>_D$ over the domain $D = [-w_\mathrm{V}/2, w_\mathrm{V}/2] \times [-y_\mathrm{max}, y_\mathrm{max}]$, as depicted in Fig. \ref{fig:ane_sha_fem}\,(f). The parameter $y_\mathrm{max}$ was chosen large enough to ensure that no significant current flows beyond $\pm y_\mathrm{max}$ and was kept fixed for all values of $w_\mathrm{V}$. The normalized average is shown as a function of the aspect ratio $a$ in Fig. \ref{fig:ane_sha_fem}\,(a) and (b). The comparison with the experimental data confirms the expected behaviour of $\theta_\mathrm{SH}^\mathrm{app} \propto \left< j_x^2 \right >_D$ and  $\alpha' \propto \left< j_x^2 \right >_D$. However, in the latter case, an offset between the measurements and the FEM simulation is seen, which might indicate a still incomplete magnetic saturation at the magnetic fields used for our measurements. We note that $V_\omega$ and, accordingly, $R_\mathrm{A}$ were found to be independent of $a$ as expected. This was also confirmed in the finite-element simulations, where we found $ \left< j_x \right >_D = \mathrm{const}$.

\section{Conclusion}
Harmonic Hall analysis experiments were performed with Hall bars with various aspect ratios. A strong dependence of the apparent spin Hall angle on the aspect ratio was observed, which was traced back to the inhomogeneity of the current and the fact that the measured second harmonic voltage depends quadratically on the $x$-component of the current density. A Hall cross with fourfold symmetry has an apparent spin Hall angle that is only about $70\%$ of the true value. The large scatter in spin Hall angles reported by different groups for nominally identical materials may to some extent be assigned to inconsistent usage of Hall bars with different aspect ratios. Thus, the aspect ratio should always be specified when reporting on harmonic Hall measurements. For an accurate determination of the spin Hall angle using the harmonic Hall measurements, Hall bars with a small aspect ratio should be preferred or the results should be corrected for the current inhomogeneity as demonstrated in the present study.

\begin{acknowledgments}
The authors thank G. Reiss for making available the laboratory equipment. They further thank T. Matalla-Wagner for support with the electron-beam lithography and for support with the construction of the vector magnet. Finally, they thank J. Balluff for providing a Python extension for the communication with the DAC.
\end{acknowledgments}


\begin{thebibliography}{50}
\bibitem{Note} We note that the equations for the harmonic Hall resistance in Ref. \onlinecite{Avci2014b} suggest that $V_{2\omega, \mathrm{ANE}} \propto I_0^2 \nabla T$, which is not correct. Here we rewrite the equations in the correct form.
\end{thebibliography}
\end{document}